\documentclass[12pt]{article}
\usepackage{graphicx}
\usepackage[cp1251]{inputenc}  
\usepackage[russian]{babel}
\usepackage{latexsym}
\usepackage{amssymb}
\usepackage{amsmath}

\begin{document}

\makeatletter
\renewcommand*{\@cite}[2]{{#2}}
\renewcommand*{\@biblabel}[1]{#1.\hfill}
\makeatother

\title{On the Properties of the Galactic Dust Layer within 700 pc of the Sun}
\author{G.~A.~Gontcharov\thanks{E-mail: georgegontcharov@yahoo.com}}
\date{Pulkovo Astronomical Observatory, Russian Academy of Sciences, Pulkovskoe sh. 65, St. Petersburg, 196140 Russia}

\maketitle

\newpage

ABSTRACT
We compare the spatial stellar color variations with our three-dimensional analytical model of the spatial dust distribution to 
refine the properties of the dust layer in Galactic solar neighborhoods. We use a complete sample of 93\,992 clump giants with a 
small admixture of branch giants from the Gaia DR2 catalogue in a spatial cylinder with a radius of 700 pc around the Sun 
extending to $|Z|=1800$ pc along the Galactic $Z$ axis. Accurate parallaxes and photometry of these stars in the Gaia DR2 
$G_\mathrm{RP}$ and WISE $W3$ bands have allowed the spatial $G_\mathrm{RP}-W3$ color variations to be used to calculate the 
model parameters and two characteristics of the sample, the mode of the dereddened color $(G_\mathrm{RP}-W3)_0$ of the giant clump
and the linear change of this mode with coordinate $|Z|$. As a result, an improved version of the three-dimensional model
first proposed by Gontcharov (2009b) has been obtained. As in the previous version, the model suggests two dust layers, 
along the Galactic equator and in the Gould Belt, that intersect near the Sun at an angle of 18$^{\circ}\pm2^{\circ}$. 
In contrast to the previous version of the model with a midplane of the Gould Belt dust layer in the form of a circle with 
the center at the Sun, in the new version this midplane is an ellipse decentered relative to the Sun. A scale height of 
$170\pm40$ pc has been found for both dust layers. A rather large reddening $E(G_\mathrm{RP}-W3)=0.16\pm0.02$ through half of 
the Galactic dust layer above or below the Sun has been found for giants far from the Galactic plane ($|Z|>600$ pc). 
This can be explained by a possible difference between the extinction law far from the Galactic plane and the commonly adopted 
law by Cardelli et al. (1989) with $R_\mathrm{V}=3.1$. The modes of the absolute magnitude $M_\mathrm{W3}=-1.70\pm0.02$ and the
dereddened color $(G_\mathrm{RP}-W3)_0=(1.43\pm0.01)-(0.020\pm0.007)|Z|$, where $Z$ is expressed in kpc, have been calculated for 
the giant clump near the Sun. These estimates are consistent with the estimates from the theoretical PARSEC and MIST isochrones 
for a sample dominated by giants with an age of 2 Gyr and metallicity $[Fe/H]=-0.1$ in agreement with the TRILEGAL stellar 
population model. The dispersions of the quantities under consideration have allowed the natural small-scale density fluctuations 
of the dust medium relative to the mean reddening calculated from the model to be characterized. These fluctuations make a major 
contribution to the uncertainty in the reddening. Because of them, the reddening of a specific star can differ from the model 
reddening by a random value that decreases from 80 to $<$20\% of the model reddening when passing from low latitudes far from 
the Sun to the remaining space.
\bigskip\noindent
\leftline {PACS numbers: 97.10.Zr; 98.35.Pr}
\bigskip\noindent
{\it Keywords:}  Hertzsprung-Russell, color-magnitude, and color-color diagrams; Solar neighborhood in Galaxy; giant stars.
\bigskip

\newpage

\section*{INTRODUCTION}

The properties of the dust layer in our Galaxy are usually determined from the observations of stars within or behind this layer. 
Therefore, until now the errors in the distances and photometry of stars have limited our knowledge of this layer.

For example, the Hipparcos parallaxes are sufficiently accurate only within $\sim100$ pc of the Sun, while the dust layer extends 
along the Galactic coordinate $Z$ (toward the Galactic Poles) and, a fortiori, in other directions much farther. For example, the
simplest dust distribution model that is very popular among researchers suggests an exponential dust distribution in $|Z|$ in one 
layer and, consequently, a change in the cumulative reddening according to the barometric law (Parenago 1954, p. 265)
\begin{equation}
\label{baro}
E(B-V)_\mathrm{R}=E(B-V)\,(1-\mathrm{e}^{-|Z-Z_0|/Z_\mathrm{A}})\,,
\end{equation}
where $E(B-V)_\mathrm{R}$ is the reddening to a distance $R$, $E(B-V)$ is the reddening to infinity on the same line of sight, 
$Z_0$ is the shift of the dust layer midplane along $Z$ relative to the Sun, and $Z_\mathrm{A}$ is the scale height of the dust 
layer. In that case, the thickness of the layer containing, say, 95\% of the dust and, thus, making an appreciable contribution 
to the reddening and extinction estimates exceeds $Z_\mathrm{A}$ by a factor of 3. Various estimates of the dust layer scale 
height are given, for example, in Perryman (2009, pp. 470--471, 496--497). Examples of the variety of such estimates for the dust 
layer near the Sun are: $<70$ pc (Juric et al. 2008), $\approx100$ pc (Gontcharov 2012b), 140 pc (Robin et al. 2003), and 
188 pc (Drimmel and Spergel 2001). Thus, assuming $Z_\mathrm{A}<200$ pc, we should use the photometry for a complete sample of
stars at least up to $|Z|<600$ pc to trace the density variations of the dust medium across the layer.

The results of the Gaia project, including Gaia DR2 (Gaia 2018a), allow one to look so far for the first time. The parallaxes 
free from significant systematic and random errors (a relative parallax accuracy of 10\% is enough for this) and the photometry
in the Gaia $G$, $G_\mathrm{BP}$ and $G_\mathrm{RP}$ bands with a median accuracy better than 0.01 mag are now known at
least within 2 kpc of the Sun for all O--G main sequence stars and all giants (Gaia 2018b). This allows complete samples of stars 
of various types in wide Galactic solar neighborhoods both inside and outside the dust layer to be analyzed for the first time.
As a result, the key characteristics of this layer can be determined. This study is the first study of such a kind.

The real dust distribution in Galactic solar neighborhoods is probably more complex than the exponential distribution in a 
single layer. This is suggested, in particular, by the following studies.
\begin{itemize}
\item According to Drimmel and Spergel (2001), the scale height of the dust layer changes with Galactocentric distance. Therefore, 
in our first study it is worth restricting ourselves to a space where the characteristics of the dust layer do not change or 
change insignificantly.
\item Kos et al. (2014) obtained estimates of the dust layer scale height for different wavelengths of the dust-absorbed radiation 
that differ approximately by a factor of 2: $118\pm5$ and $209\pm12$ pc, respectively, for the best studied dust fraction causing 
extinction in the visible range, at a wavelength of about 0.5 microns, and for the dust producing the diffuse interstellar
line at 0.862 microns. In the presence of a natural relationship between the size of a dust grain and the wavelength of the 
radiation absorbed and scattered by it (for a review, see Gontcharov 2016b), the result by Kos et al. (2014) suggests that 
larger dust grains produce a thicker layer in the Galaxy. One consequence of this would be the dependence of the extinction law 
(i.e., the extinction as a function of wavelength) on $|Z|$. Indeed, such a dependence was found in several studies 
(Gontcharov 2016b) and, in particular, by Gontcharov (2012a, 2013b, 2016a), in studies similar to this one, when analyzing complete
samples of giants in regions of space extended in $Z$. Therefore, it is worth assuming that the dust layer characteristics depend 
on the stellar color under consideration and, consequently, considering different colors independently. In this study we consider 
only one color. This does not allow any conclusions about the extinction law in the space under consideration to be reached, but, 
on the other hand, the inferred dust layer characteristics do not depend on the extinction law.
\item As the numerous tests mentioned below show, the model of the dust distribution in two layers, the equatorial (near the
Galactic midplane) one and the layer in the Gould Belt tilted to it, proposed by Gontcharov (2009b, 2012b) corresponds to the 
observations better than does the model with an exponential dust distribution in one layer for any characteristics of this layer 
and better than do the previous models by Arenou et al. (1992) and Drimmel and Spergel (2001) (implemented in the three-dimensional 
reddening map by Drimmel et al. (2003)). Therefore, the model by Gontcharov (2009b, 2012b), its new, more realistic version is 
considered in this study. The Gould Belt was apparently first considered as a dust container and a cause of the stellar reddening 
by Vergely et al. (1998). However, the final conclusion about the existence of an additional dust layer in the Gould Belt can be
reached only on the basis of several studies using various present-day data (different three-dimensional reddening maps, photometry 
from different surveys).
\end{itemize}

The parameters of the model under consideration were calculated in the first (Gontcharov 2009b) and second (Gontcharov 2012b) 
solutions using various reddening and extinction data sources.

In this study the characteristics of the Galactic dust layer in solar neighborhoods are calculated within the new version of 
the model under consideration by analyzing the spatial variations of the observed color for a complete sample of giants from 
Gaia DR2 in combination with two characteristics of these giants, their dereddened color near the Sun and the change of the 
dereddened color with coordinate $|Z|$. Some of the characteristics determined in this study (for example, the total reddening 
through the dust half-layer toward the Galactic Poles, the dereddened color of stars near the Sun, and the change of the
dereddened color with coordinate $|Z|$) do not depend on the adopted model, because they are actually calculated from the color 
difference for giants close to the Sun and giants of the same type that are definitely outside the dust layer.

Any reddening/extinction model has an insurmountable limitation. The model assigns the reddening/extinction averaged for some 
region of space around a point of space to this point. However, small-scale density fluctuations of the dust medium exist within 
even the smallest averaging region. In any case, they manifest themselves on the scale of a typical distance between neighboring 
stars for any sample of Galactic stars, given that this distance is relatively large. According to the Besancon Galaxy
model (Czekaj et al. 2014), the spatial density of stars of all types in solar neighborhoods does not exceed 0.05 solar mass per 
cubic parsec, i.e., given the dominance of red dwarfs with a mass of $\sim0.1$ solar mass, the mean distance between stars is 
$\sim1$ pc. Therefore, any model cannot take into account these small-scale fluctuations in principle (although they can be 
taken into account when the reddening of a star is calculated from its individual data). Nevertheless, the observed scatter of 
colors allows statistical estimates of the small-scale fluctuations of the dust medium to be obtained. This subject has been hardly
investigated (see the reasoning in Green et al. 2015). However, it is obvious that the density fluctuations of the dust medium 
decrease as the density itself decreases in such a way that, given the fluctuations, the density remains nonnegative at any point 
of space. In this study we make an attempt to estimate these fluctuations and the corresponding constraints on the applicability 
of the model under consideration based on the observed scatter of colors.

\section*{MODEL SELECTION}

A lot of maps and analytical models have been constructed to describe the spatial distribution of dust, the corresponding 
reddening of stars, and interstellar extinction. The difference between them is that the maps and models describe the dust medium
at a point of space by a particular number and a particular function of Galactic coordinates, respectively. In contrast to the 
maps, the models explicitly or implicitly include the dust layer characteristics.

The real accuracy (primarily in systematic terms) of the most popular and accurate models and maps has been analyzed in recent 
years using accurate Gaia parallaxes, theoretical isochrones, and Galaxy models.
\begin{enumerate}
\item Gontcharov (2017a) and Gontcharov and Mosenkov (2017a, 2017b, 2018) placed stars with accurate parallaxes from the Gaia DR1
TGAS catalogue (Michalik et al. 2015) and accurate photometry from various sky surveys on the Hertzsprung–Russell (HR) diagram to
analyze the distribution of these stars relative to the theoretical PAdova and TRieste Stellar Evolution Code (PARSEC) 
(Bressan et al. 2012; http://stev.oapd.inaf.it/cmd), MESA Isochrones and Stellar Tracks (MIST) (Paxton et al. 2011; Choi et al. 
2016; Dotter 2016; http://waps.cfa.harvard.edu/MIST/), Yale--Potsdam Stellar Isochrones (YaPSI) (Spada et al. 2017; 
http://www.astro.yale.edu/yapsi/) isochrones and the theoretical distributions from the Besancon Galaxy model
and the TRILEGAL Galaxy model (Girardi et al. 2005; http://stev.oapd.inaf.it/cgi-bin/trilegal).
\item Gontcharov and Mosenkov (2019) used the estimates of the reddening $E(B-V)$ and interstellar polarization $P$ for Gaia DR2 
stars within 500 pc of the Sun to compare the spatial variations of the effective polarization $P/E(B-V)$ with the theoretical 
views of the interstellar medium. Furthermore, Gontcharov and Mosenkov (2019) reprocessed the data from Welsh et al. (2010) on 
the spatial variations of the equivalent widths of Na I and Ca II absorption lines, which are indicative of the three-dimensional 
interstellar gas and dust distribution, using the Gaia DR2 parallaxes and compared them with the estimates based on various 
reddening models and maps.
\item Gontcharov et al. (2019) analyzed the distribution of stars in the Galactic globular cluster NGC 5904 (M5) on the 
color–magnitude diagram based on 29-band photometry relative to the theoretical PARSEC, MIST, Dartmouth Stellar Evolution Program 
(DSEP) (Dotter et al. 2007; http://stellar.dartmouth.edu/models/) and A Bag of Stellar Tracks and Isochrones (BaSTI) 
(Pietrinferni et al. 2004; http:// basti.oa-teramo.inaf.it) isochrones and calculated the most probable characteristics
of this cluster, including the extinction in all 29 bands, the corresponding reddenings, and the extinction law.
\end{enumerate} 

These and other tests have shown that the reddening models and maps that are based on the data for complete samples of stars and 
take into account the nonuniform dust distribution near the Sun (at a distance $R<400$ pc), including the variations of the
dust distribution with Galactic longitude $l$, are most accurate. In particular, the three-dimensional maps by Green et al. 
(2015, 2018), which are most accurate far from the Sun, show a low accuracy near the Sun, because they are based on the photometry 
of distant stars, while for the nearest $\sim400$ pc they simply interpolate the reddening between the reddening for these
stars and zero reddening near the Sun. It is worth mentioning the three-dimensional map by Lallement et al. (2018) as a map 
showing clear systematic errors due to significant incompleteness of the sample of stars used.

In the mentioned tests the model with an exponential dust distribution in one layer was applied to the best two-dimensional 
(i.e., with estimates of the reddening through the entire Galactic dust half-layer from the Sun to infinity) reddening
maps by Schlegel et al. (1998) and Meisner and Finkbeiner (2015). These maps were constructed from the data on the infrared (IR) 
dust emission using the IRAS/ISSA + COBE/DIRBE and IRAS/ISSA + Planck telescopes, respectively, and were calibrated based on 
the reddening estimates for elliptical galaxies, quasars, and stars. The tests showed that the exponential dust distribution is too
far from reality, because it gives erroneous (primarily in systematic terms) reddening estimates for real stars.

Among the remaining three models of the dust distribution in Galactic solar neighborhoods, the model by Gontcharov (2009b, 2012b) 
showed the best results in the tests. The explicit allowance for an additional dust layer in the Gould Belt and the variation of 
the spatial dust density with longitude in both layers, the equatorial one and the Gold Belt layer, are apparently its advantages. 
The models by Arenou et al. (1992) and Drimmel and Spergel (2001) turned out to be worse. The reasons for this were actually 
pointed out by the authors of these models. The model by Arenou et al. (1992) relies on the observations of a relatively small 
number of stars, especially far from the Galactic plane, and, as a result, this model has to restrict itself to constant 
extinctions for vast high-latitude sky regions. However, at low latitudes the model by Arenou et al. (1992) showed good results 
and extinctions close to the values from the model by Gontcharov (2009b, 2012b). The model by Drimmel and Spergel (2001) takes 
into account the nonuniformly distributed dust in Galactic solar neighborhoods in the form of an Orion--Cygnus arm segment. 
However, the authors themselves recognize that ideally this region about the Sun should be described by a more detailed local 
model of the dust distribution (Drimmel et al. 2003).

\section*{MODEL DESCRIPTION}

The model under consideration describes the spatial reddening variations within several hundred parsecs of the Sun, i.e., in a 
small part of the Galaxy. Therefore, the equatorial dust layer considered in the model may be deemed infinite in the $X$ and $Y$ 
directions. In contrast to it, the dust layer in the Gould Belt has a finite size. In the previous versions of the model
it had a circular midplane with the center at the Sun. Now this midplane is an ellipse with a center shifted relative to the Sun 
in all three coordinates, a semimajor axis $A$, a semiminor axis $a$, an eccentricity $e$, and an angle $\eta$ between the 
semimajor axis and the direction of maximum reddening. The reddening in the Belt is calculated toward a star only to a distance
$R_0$ from the Sun to the Belt edge if the star is farther than the edge or to the star if it is nearer than the edge.
The stellar coordinates in the Gould Belt coordinate system are: $\zeta$ is the shortest distance from the star to
the Belt midplane (an analogue of $Z$ in the Galactic coordinate system), $\beta$ is the latitude measured from the Belt midplane, 
and $\lambda$ is the longitude measured from the direction of maximum reddening in the Belt layer. The layers intersect at an 
angle $\gamma$. The angle between the $Y$ coordinate axis and the line of intersection of the layers is denoted by $\theta$. 
These quantities are related by the following relations:
\begin{equation}
\label{ellipse}
a^2=A^2(1-e^2)
\end{equation}
\begin{equation}
\label{r0}
R_0=a/(1-(e\cos(\lambda-\eta))^2)^{1/2}
\end{equation}
\begin{equation}
\label{zeta}
\zeta=\min(R,R_0)\sin(\beta)
\end{equation}
\begin{equation}
\label{beta}
\sin(\beta)=\cos(\gamma)\sin(b)-\sin(\gamma)\cos(b)\cos(l)
\end{equation}
\begin{equation}
\label{lambda}
\tg(\lambda-\theta)=\cos(b)\sin(l)/(\sin(\gamma)\sin(b)+\cos(\gamma)\cos(b)\cos(l)).
\end{equation}
The equatorial layer is shifted relative to the Sun along $Z$ by a distance $Z_\mathrm{0}$, while the Gould Belt layer
is shifted along $X$, $Y$, and $Z$ by $x_\mathrm{0}$, $y_\mathrm{0}$ and $z_\mathrm{0}$, respectively.

\begin{figure*}
\includegraphics{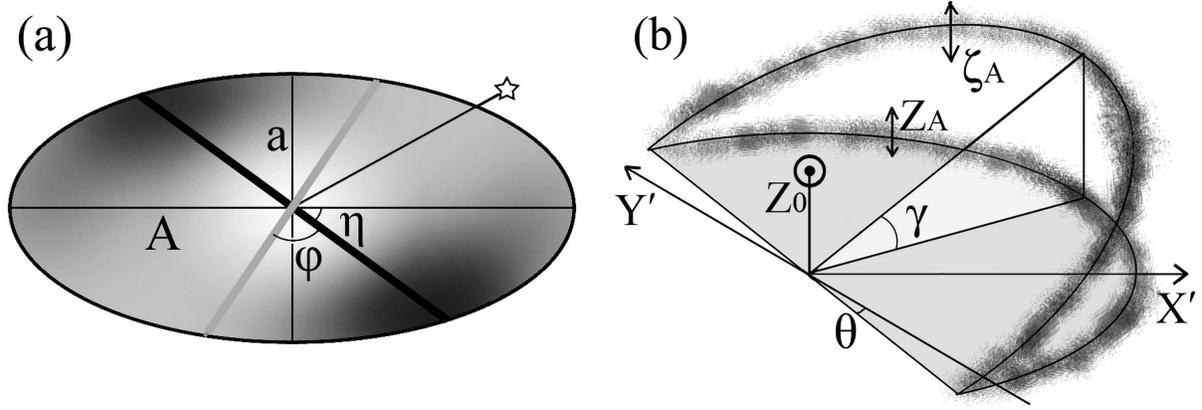}
\caption{Scheme of the elliptical dust layer midplane in the Gould Belt (a) and the intersecting midplanes of the two layers (b).
(a) The shading indicates the density of the dust distribution, the thick black line indicates the direction of maximum reddening,
the thin black lines mark the semimajor $A$ and semiminor $a$ axes of the layer ellipse, the angles $\eta$ and $\phi$ are marked, 
the thick gray line is the line of intersection of the layers (on the scheme, for simplicity, it passes through the center of 
the layer ellipse, although in the model it is shifted relative to the latter), the direction to some star is also marked. 
(b) The directions $X'$ and $Y'$ parallel to the $X$ and $Y$ axes, the Sun's elevation above the equatorial layer $Z_\mathrm{0}$, 
the scale heights $Z_A$ and $\zeta_A$, and the angles $\gamma$ and $\theta$ are marked.
}
\label{g19}
\end{figure*}

The midplanes of the dust layers in the model under consideration are schematically shown in Fig. 1: separately the elliptical 
midplane of the dust layer in the Gould Belt (Fig. 1a) and the intersecting midplanes of the two layers (Fig. 1b).

The reddening of a star is considered as a sum of the reddenings in the two layers. Each reddening is a function of Galactic 
coordinates and is described by a barometric law with sinusoidal variations as a function of longitude inside the layer:
\begin{equation}
\label{equ1}
(E0_\mathrm{equator}+E1_\mathrm{equator}\sin(l+\Phi))R(1-e^{-|Z-Z_\mathrm{0}|/Z_\mathrm{A}})Z_\mathrm{A}/|Z-Z_\mathrm{0}|
\end{equation}
for the equatorial layer and
\begin{equation}
\label{equ2}
(E0_\mathrm{Gould}+E1_\mathrm{Gould}\sin(2\lambda+\phi))\min(R,R_0)(1-e^{-|\zeta|/\zeta_\mathrm{A}})\zeta_\mathrm{A}/|\zeta|
\end{equation}
for the layer in the Gould Belt. Here, $Z_\mathrm{A}$, $E0_\mathrm{equator}$, $E1_\mathrm{equator}$ and $\Phi$ are, respectively, 
the scale height, the free term, and the reddening amplitude and phase in the sinusoidal dependence on $l$ for the equatorial
layers; $\zeta_\mathrm{A}$, $E0_\mathrm{Gould}$, $E1_\mathrm{Gould}$ and $\phi$ are, respectively, the scale height, the free term, 
and the reddening amplitude and phase in the sinusoidal dependence on $2\lambda$ for the layer in the Gould Belt. The calculations
were also performed with the term $\sin(\lambda+\phi)$ in Eq. (8), but the model with the term $\sin(2\lambda+\phi)$, i.e., 
with two reddening maxima in the Gould Belt, corresponds better to the observations. These maxima occur approximately at the 
longitudes of the dust-richest Scorpius-–Ophiuchus–-Sagittarius and Perseus–-Taurus–-Orion cloud complexes located approximately
at the longitudes of the Galactic center and anticenter, respectively. This is consistent with the universally accepted views 
of the dust distribution in the Gould Belt (Perryman 2009, pp. 324--328; Bobylev 2014).

\section*{INITIAL DATA}

Clump giants are most suitable for this study, because they are bright, numerous, and are relatively easily identified on the 
HR diagram based on photometry. These are stars after the passage of the giant branch and the helium flash. A clump giant
consists of an inert hydrogen envelope and a helium core, in which the nuclear helium-to-carbon fusion proceed. An overview of 
the current views of such stars is given in Girardi (2016). Samples of clump giants from various photometric surveys have been
produced in recent years by various authors (for a review, see Gontcharov 2016b) and have already been used, among other things, 
to analyze the properties of the dust medium in wide Galactic solar neighborhoods (Gontcharov et al. 2013a, 2013b, 2016a).

As has been noted in the Introduction, the accuracy of the Gaia DR2 parallaxes and photometry allows a complete sample of 
clump giants within at least 2 kpc of the Sun to be examined. To test the hypothesis about reddening in the Gould Belt
dust layer, we will restrict ourselves to the space where the Belt must play an important role. The Belt resembles a relatively 
flat disk with a radius of $\sim400-500$ pc decentered relative to the Sun by several tens of pc and tilted to the Galactic 
plane at an angle of about 20$^{\circ}$ (Perryman 2009, pp. 324--328; Bobylev 2014). Therefore, let us introduce the constraint
$(X^2+Y^2)^{1/2}<700$ pc, where $X$ and $Y$ are the Galactic rectangular coordinates in the direction of the Galactic center 
and its rotation, respectively. In addition, given the drop in the density of the distribution of stars with $|Z|$, we will limit 
the sample to $|Z|<1800$ pc to avoid the unjustified influence of a few stars with large |Z| on the result. Thus, we consider a 
spatial cylinder with a radius of 700 pc around the Sun extending to $|Z|<1800$ pc.

The distances derived by Bailer-Jones et al. (2018) from the Gaia DR2 parallaxes were used for the sample. The errors of these 
distances are so small that they do not affect the results of this study: the median of the relative distance error is 2\%. 
The distances of 658 sample stars (0.7\%) have a relative accuracy worse than 10\%, but these stars were left in the sample, 
because they are distributed quite uniformly in space and do not affect the results.

The set of bands with accurate (the median accuracy is 0.01 mag) photometry for clump giants in the entire space under 
consideration is very limited (for an analysis, see Gontcharov 2016a). For example, given the absolute magnitudes of clump 
giants, their photometry fromthe Tycho-2 catalogue (Hoeg et al. 2000) is accurate and, consequently, their sample is complete
only to 600 and 740 pc in the $B_\mathrm{T}$ and $V_\mathrm{T}$ bands, respectively. In contrast, the IR photometry of clump giants is generally 
inaccurate for the stars closest to the Sun, because they are too bright in the IR for many detectors. For example, in the $J$, $H$, 
and $Ks$ bands of the Two Micron All-Sky Survey (2MASS) catalogue (Skrutskie et al. 2006) and the $W1$ and $W2$ bands of the 
allWISE catalogue (Wright et al. 2010; http://irsa.ipac.caltech.edu/Missions/wise.html) of the Wide-field Infrared Survey Explorer 
(WISE) telescope the photometry of clump giants is inaccurate within 150--500 pc of the Sun (depending on the band), i.e., in 
the region with maximum variations in the density of the dust distribution most important for us.

Gaia DR2 is the first survey with accurate optical photometry for clump giants in the entire space under consideration. Therefore, 
to obtain a preliminary complete sample, we used a $(G_\mathrm{BP}-G_\mathrm{RP})$ versus $(G_\mathrm{RP}+5-5\log(R))$ HR diagram. 
We selected all stars in the giant clump according to the criteria 
$1.0<G_\mathrm{BP}-G_\mathrm{RP}<2.3$, $-0.6<G_\mathrm{RP}+5-5\log(R)<1.5$ and the additional criteria for the elimination 
of subgiants
$G+5-5\log(R)<1.75(G_\mathrm{BP}-G_\mathrm{RP})-1.25$, $G+5-5\log(R)>1.92(G_\mathrm{BP}-G_\mathrm{RP})-2.12$, without
any correction for reddening and extinction. The photometry of the selected stars turned out to be very
accurate: for all stars $\sigma(G_\mathrm{BP})<0.05$, $\sigma(G_\mathrm{RP})<0.03$, the medians $\sigma(G_\mathrm{BP})$ and 
$\sigma(G_\mathrm{RP})$ are 0.001.

For more reliable results it is worth using an IR band in combination with an optical band. Almost all stars in the preliminary 
sample have accurate photometry in the allWISE $W3$ band (with an effective wavelength of 10.8 microns). The $W3$ band has already
been used in a similar study by Gontcharov (2017), who showed that it is convenient, among other things, owing to low extinction. 
The final sample was formed from the preliminary one by selection according to the criteria $G_\mathrm{RP}-W3>1$ and 
$-2<W3+5-5\log(R)<-1.25$ without any correction for reddening and extinction. In addition, 94 stars with inaccurate $W3$ photometry 
were rejected. As a result, the sample contains 93\,992 stars.

Having lost only 94 stars with inaccurate photometry in the $W3$ band, we note, for comparison, that 1637, 10\,328, 1304, 40\,609, 
7489, and 77\,355 sample stars have no photometry more accurate than 0.05 mag in the IR 2MASS $J$, $H$, $Ks$ and allWISE
$W1$, $W2$, $W4$ bands, respectively. Consequently, only using the $W3$ band maintains the sample completeness at an acceptable 
level. Apparently, the $J$ and $Ks$ bands can also be used in future studies after an additional study of how much the corresponding
incompleteness affects the result. Obviously, the remaining bands cannot be used in such a study.

\begin{figure*}
\includegraphics{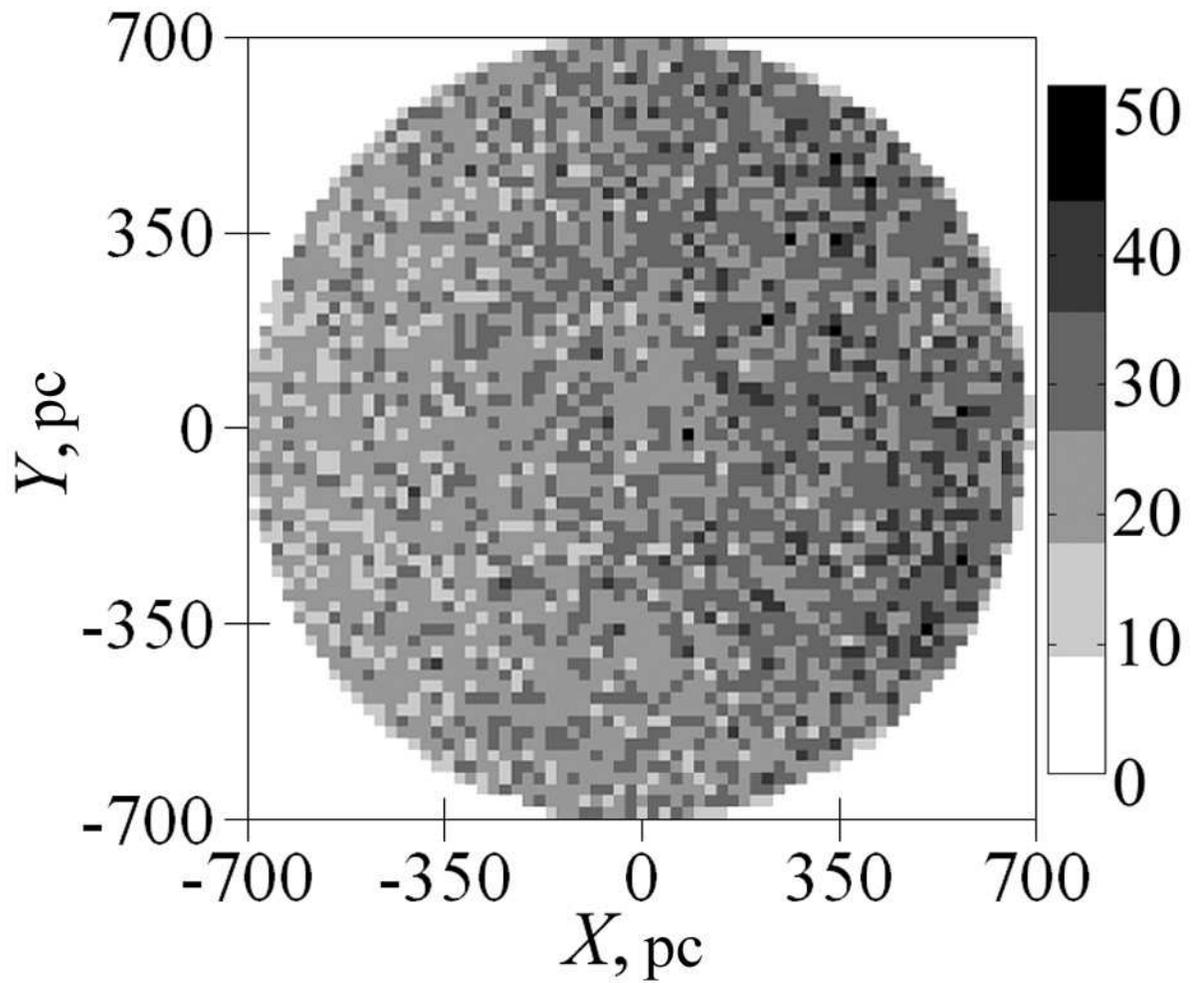}
\caption{Distribution of the sample stars in projection onto the $XY$ plane in $20 \times 20$ pc squares. The Sun is at the center, 
the Galactic center is on the right. The shading scale is given on the right.
}
\label{xy}
\end{figure*}

The completeness of our sample is seen in Fig. 2, where the distribution of the sample stars in projection onto the $XY$ plane in 
$20 \times 20$ pc squares is shown. As expected, the spatial distribution of the stars increases weakly toward the Galactic center
(on the right), but does not decrease with distance from the Sun, which would be in the case of sample incompleteness.

The results of this study could be affected by the very large width of the $G$ band and the large separation between the 
$G_\mathrm{BP}$ and $W3$ bands. Therefore, here we consider only one of the possible colors, $G_\mathrm{RP}-W3$.

Figure 3 shows the distribution of the sample stars on a $(G_\mathrm{RP}-W3)$ -- $(M_\mathrm{W3})$ HR diagram before (a) and
after (b) the correction for the reddening $E(G_\mathrm{RP}-W3)$ found in this study and for the extinction $A_\mathrm{W3}$
calculated as
\begin{equation}
\label{reddeningextinction}
A_\mathrm{W3}=0.0135E(G_\mathrm{RP}-W3)
\end{equation}
in accordance with the extinction law by Cardelli et al. (1989, hereafter CCM89) with $A_\mathrm{V}/E(B-V)=3.1$. 
This law also suggests $A_\mathrm{G_{RP}}=2.03E(B-V)$, $A_\mathrm{W3}=0.027E(B-V)$ and $E(G_\mathrm{RP}-W3)=2.00E(B-V)$. 
Note that the reddening $E(G_\mathrm{RP}-W3)$ was calculated based exclusively on the observed spatial variations of the 
$G_\mathrm{RP}-W3$ color and, therefore, is completely independent of the extinction law. Moreover, using a specific extinction 
law for this HR diagram changes its shape insignificantly, because the coefficient in relation (9) is very small.
Note also that the CCM89 extinction law can in no way be considered as some `standard', especially in the IR, because, for 
example, the extinction law by Weingartner and Draine (2001), which is equally justified by observations, gives the relation
\begin{equation}
\label{wd2001coef}
A_\mathrm{W3}=0.14E(G_\mathrm{RP}-W3),
\end{equation}
with a coefficient that is greater than the coefficient in relation (9) by an order of magnitude.

\begin{figure*}
\includegraphics{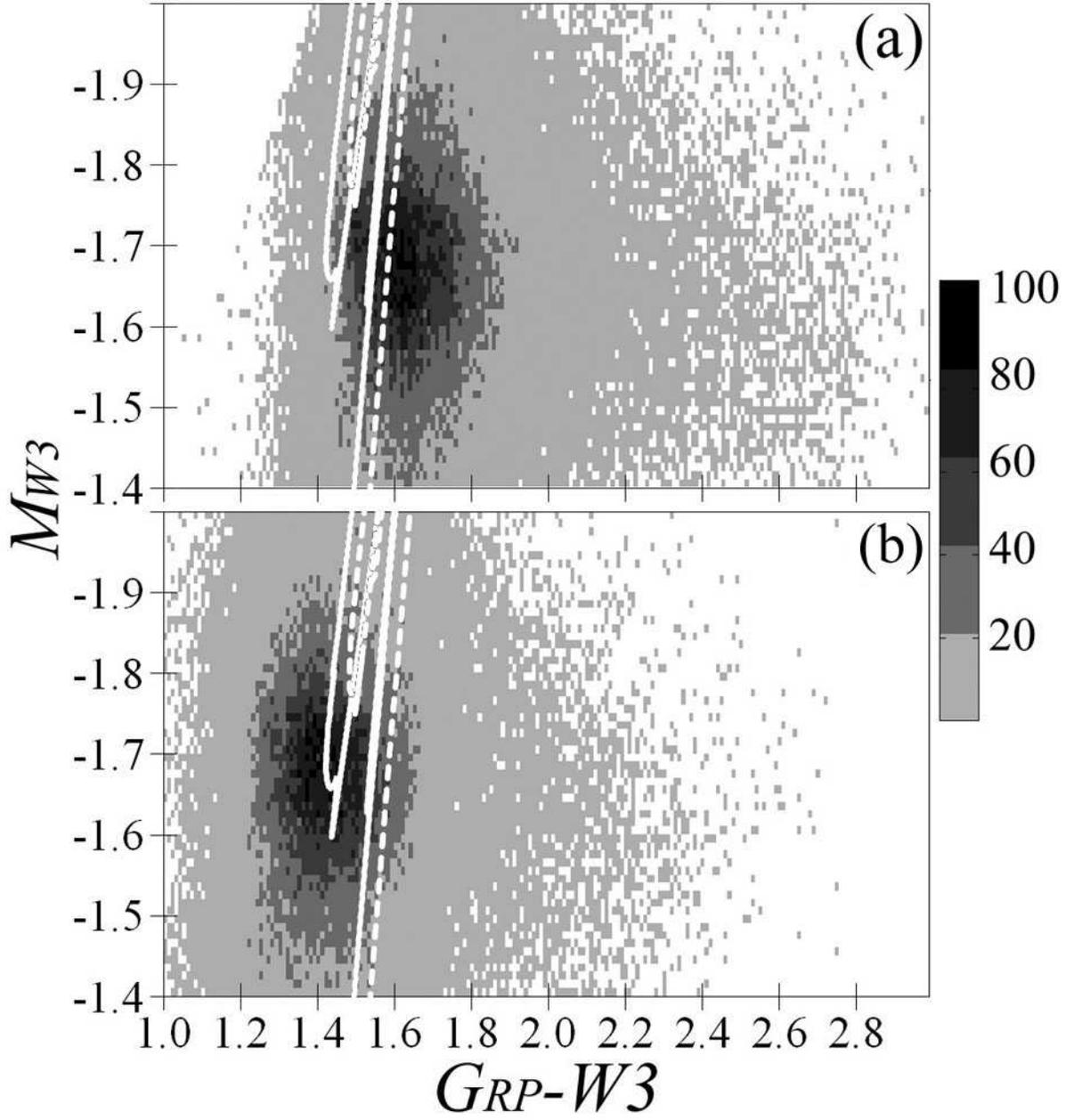}
\caption{$(G_\mathrm{RP}-W3)$ -- $M_\mathrm{W3}$ HR diagram for the sample stars before (a) and after (b) the correction for 
reddening $E(G_\mathrm{RP}-W3)$ and extinction $A_\mathrm{W3}$. The PARSEC and MIST isochrones for an age of 2 Gyr and 
metallicity $[Fe/H]=-0.1$ are indicated by the solid and dashed white curves, respectively. The shading scale for the 
distribution of stars in $0.01 \times 0.01$ mag cells is given on the right.
}
\label{hr}
\end{figure*}

For the sample stars the median of $|Z|$ is 212 pc; 74 and 90\% of the stars have $|Z|<400$ and 700 pc, respectively. Consequently, 
most of the sample stars are within the Galactic dust layer. The distribution of the sample along $Z$ is nearly normal with a 
mean of $-22$ pc (the shift of the Sun along $Z$ relative to the Galactic midplane manifested itself in this way) and
a standard deviation of 426 pc. This distribution is typical for the clump giants in solar neighborhoods (Gontcharov 2008, Fig. 6). 
Such a spatial distribution of the sample is convenient for studies of the dust layer. However, we cannot reach any conclusions
about the spatial distribution of precisely the clump giants, because in this study we did not apply any methods of removing the 
admixture of branch giants from the sample considered, for example, by Gontcharov (2008, 2009a). The branch giants in the
same space have characteristics (dereddened color, absolute magnitude, age, and metallicity) slightly different from those of 
the clump giants. However, using a sample of clump giants from Gaia DR1 as an example, Gontcharov (2017) showed that an admixture
of branch giants in this sample affects significantly the spatial variations of the mean $G_\mathrm{RP}-W3$ and 
$(G_\mathrm{RP}-W3)_0$, but does not affect the modes of these colors at all, because the clump giants constitute an
overwhelming majority everywhere in the region of theHR diagram under consideration and the region of space under consideration. 
To estimate the influence of an admixture of branch giants on the result in this study, we consider both the mean and the mode of
$G_\mathrm{RP}-W3$. The mode was calculated for cells of space containing 400 stars each.

\section*{RESULTS}

In this study the spatial $G_\mathrm{RP}-W3$ color variations are considered as a result of the following: (1)
the reddening of stars and (2) the systematic linear change $\Delta(G_\mathrm{RP}-W3)_0$ of their dereddened color
$(G_\mathrm{RP}-W3)_0$ with coordinate $|Z|$ due to the change in their age and metallicity.

The solution for 19 sought-for parameters is presented in Table 1. This set of parameters corresponds to the minimum sum of the 
squares of the residuals, i.e., the differences between the model and observed colors $G_\mathrm{RP}-W3$.

Based on the existing views of the dust distribution in Galactic solar neighborhoods and the Gould Belt and taking into account 
the previous solutions, for each sought-for parameter we specified the range of acceptable values given in Table 1 
(for convenience, both characteristics of the ellipse $a$ and $e$ dependent on each other and the sum of the scale heights
$Z_A+\zeta_A$ are given). For each parameter within the range we adopted a uniform grid of five values of the parameter. 
Owing to the power of modern computers, the solution was sought by the exhaustion of values for each of the 19 sought-for 
parameters on the specified grid (i.e., the sum of the squares of the residuals was calculated for 519 sets of parameters).
As a result, for each parameter we chose two values for which the sum of the squares of the residuals was minimal. Within 
the interval between these two values we again adopted a uniform grid of 3--6 values of the parameter and the calculations 
(i.e., the second iteration) were repeated on this new grid. No more than four iterations were needed for all parameters.
Given the auxiliary calculations, we considered a total of 10$^{14}$ sets of parameters.

For some parameters the sum of the squares of the residuals for all grid values differed by no more than 0.2\% already in the 
second iteration. As the error in each parameter given in Table 1we adopted its change that gave a change in the standard deviation
of the residuals no more than 0.2\% at fixed remaining parameters. The values of the parameters found using the individual 
$G_\mathrm{RP}-W3$ and $mode(G_\mathrm{RP}-W3)$ coincided within the limits of the errors given in Table 1.

Note that some parameters were calculated with a great uncertainty, because their variation in a relatively wide range hardly 
affects the sum of the squares of the residuals. This is also pointed out in the previous solutions (Gontcharov 2009b, 2012b) and
is explained by the features of the model. For example, the angle $\gamma$ between the dust layers is small enough for the 
two layers to actually appear as one in a large region of space ($b\approx0^{\circ}$ and $l\approx\pm90^{\circ}$). 
Therefore, $\gamma$ variations in the relatively wide range $18\pm2^{\circ}$ change the sum of the squares of the residuals 
by less than 0.2\%. Furthermore, the relatively large reddening found far from the Galactic plane at a rather low reddening 
near this plane forces us to adopt relatively large scale heights for the layers in accordance with Eqs. (7) and (8). 
Since the angle $\gamma$ is small, far from the Galactic plane an increase in the scale height of each layer gives approximately 
the same effect. Therefore, only the sum of the scale heights of the layers is determined with confidence, while each scale height 
is calculated with a great uncertainty. In other words, the scale heights partly depend on each other: increasing one, we are forced
to reduce the other approximately by as much. As a result, we adopted identical scale heights for both layers.

\begin{table*}
\def\baselinestretch{1}\normalsize\normalsize
\caption[]{The ranges of model parameters under consideration and the solution found
}
\label{solution}
\[
\begin{tabular}{lcc}
\hline
\noalign{\smallskip}
 Parameter        & Range & Solution \\
\hline
\noalign{\smallskip}
$\gamma$, $^{\circ}$                            & $14-26$        & $18\pm2$       \\ 
$\theta$, $^{\circ}$                            & $-60-+60$      & $15\pm5$       \\ 
$\eta$, $^{\circ}$                              & $-90-+90$      & $28\pm5$       \\ 
$A$, pc                                         & $400-700$      & $600\pm50$     \\ 
$e$                                             & $0.34-0.98$    & $0.95\pm0.02$  \\
$a$, pc                                         & $80-668$       & $187\pm50$     \\ 
$x_0$, pc                                       & $-160-+160$       & $+20\pm30$       \\ 
$y_0$, pc                                       & $-160-+160$       & $-100\pm30$      \\ 
$z_0$, pc                                       & $-50-+50$      & $+7\pm10$      \\ 
$Z_0$, pc                                       & $-30-+30$      & $-10\pm5$       \\ 
$Z_A$, pc                                       & $70-270$       & $170\pm40$     \\ 
$\zeta_A$, pc                                   & $40-440$       & $170\pm40$     \\ 
$Z_A+\zeta_A$, pc                               & $110-710$      & $340\pm30$    \\
$E0_\mathrm{equator}$, $^m$~kpc$^{-1}$          & $0.4-1.2$      & $0.59\pm0.05$  \\ 
$E1_\mathrm{equator}$, $^m$~kpc$^{-1}$          & $0.0-0.6$      & $0.15\pm0.05$  \\ 
$\Phi$, $^{\circ}$                              & $-90-+90$      & $40\pm10$       \\ 
$E0_\mathrm{Gould}$, $^m$~kpc$^{-1}$            & $0.0-1.2$      & $0.48\pm0.05$  \\ 
$E1_\mathrm{Gould}$, $^m$~kpc$^{-1}$            & $0.0-1.2$      & $0.26\pm0.05$  \\
$\phi$, $^{\circ}$                              & $0-180$        & $120\pm15$       \\
$(G_\mathrm{RP}-W3)_0$, $^m$                    & $1.40-1.52$    & $1.43\pm0.01$  \\ 
$\Delta(G_\mathrm{RP}-W3)_0$, $^m$~kpc$^{-1}$  & $-0.08-0.00$   & $-0.020\pm0.007$ \\ 
\hline
\end{tabular}
\]
\end{table*}

The sizes of the dust layer in the Gould Belt, i.e., $A$ and $a$, were also calculated with a great uncertainty, although the 
eccentricity $e$ (i.e., the layer shape) is determined with confidence. This is because the outer boundary of the layer can be 
defined only as a region where the spatial density of the dust distribution and, accordingly, the differential interstellar 
extinction (measured in magnitudes per parsec) decreases. However, the change in the color of stars used by us as data reflects 
not the differential, but cumulative extinction (from the Sun to the star measured in magnitudes). The cumulative 
reddening/extinction should increase as one recedes from the Sun (or the layer center) and should be stabilized at the layer
boundary. However, given the natural fluctuations of the dust medium and the well-known general rise in the density of the dust 
distribution toward the Galactic center, the distance at which the cumulative reddening ceases to increase is determined with an
uncertainty. The same factors led to a great uncertainty in the coordinates $x_0$ and $y_0$ of the Gould Belt center (furthermore, 
the rise in the density of the dust distribution toward the Galactic center can lead to some shift of the $x_0$ estimate 
toward the center). The differential reddening can be calculated from the cumulative one only if the spatial density of the points
(stars) under consideration is great and there are several points (stars) on each line of sight. The density of the distribution 
of clump giants is not enough for this. However, the three-dimensional reddening maps by Gontcharov (2017a) and Lallement et al. 
(2018) have versions with differential reddening and can be used in future for a more accurate determination of the sizes
of the dust layer in the Gould Belt.

\begin{figure*}
\includegraphics{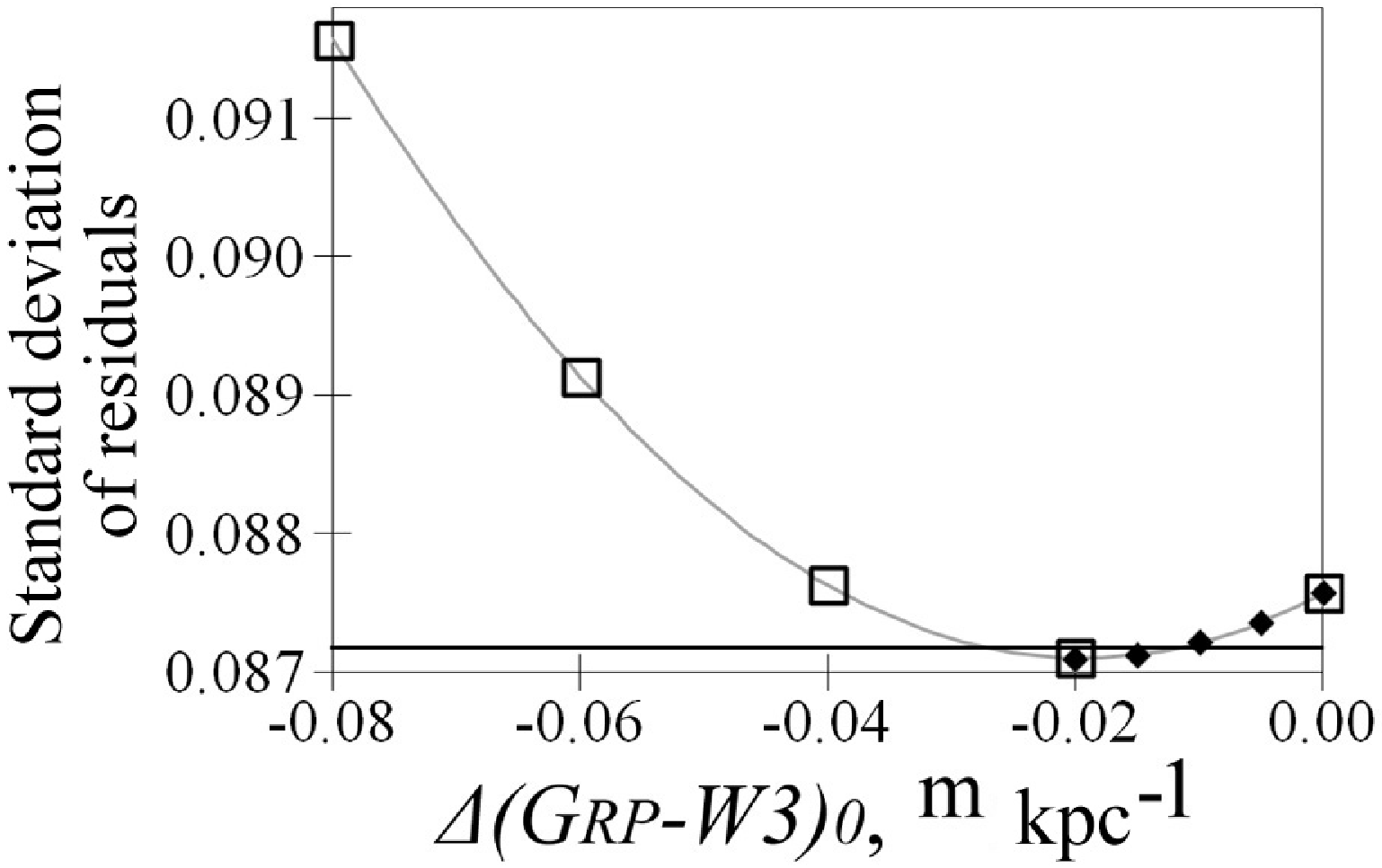}
\caption{Standard deviation of the residuals for the mode of $G_\mathrm{RP}-W3$ versus coefficient $\Delta(G_\mathrm{RP}-W3)_0$ 
at fixed values of the remaining parameters from Table 1 (gray curve). The grid of values is represented by the open squares 
in the first iteration and the filled diamonds in the second iteration. The horizontal black line marks the 0.2\% level above 
the minimum.
}
\label{metal}
\end{figure*}

The gray curve in Fig. 4 shows an example of the change in the standard deviation of the residuals for the mode of $G_\mathrm{RP}-W3$
with coefficient $\Delta(G_\mathrm{RP}-W3)_0$ at fixed values of the remaining parameters from Table 1. The horizontal black 
line marks the 0.2\% level above the minimum. Our calculations for the entire variety of the remaining parameters were
performed on the grid of $\Delta(G_\mathrm{RP}-W3)_0$ marked in Fig. 4 by the large open squares in the first iteration
and the filled diamonds in the second iteration. This is one of the cases where the solution is obvious already after the second 
iteration: $\Delta(G_\mathrm{RP}-W3)_0=-0.020\pm0.007$ mag per kpc.

We can see how successfully the model explains the spatial $G_\mathrm{RP}-W3$ variations from Figs. 5 and 6, where 
$mode(G_\mathrm{RP}-W3)$ is shown as a function of $R$ and $l$, respectively, in Fig. 6 for three layers: (a) $Z>140$, 
(b) $-140<Z<140$, (c) $Z<-140$ pc. The initial data and the model are indicated by the black squares and gray diamonds, 
respectively. The large scatter of data stems from the fact that, in this case, the modes were calculated over the entire 
range of $b$ and $l$ in Fig. 5 and over the entire range of $R$ in Fig. 6. The vertical straight line in Fig. 5 marks $R=700$ pc,
where a sharp break in the dependence is seen. It is explained by the cylindrical shape of the region of space under consideration. 
At $R<700$ and $>700$ pc the sample is dominated, respectively, by stars near the Galactic plane far from the Sun and stars far from
the Galactic plane exactly above and below the Sun. For the former allowance for their large reddenings is important, but the 
natural change of the dereddened color with $Z$ is hardly important; for the latter the reverse is true. The model well 
reproduces this break near $R=700$ pc, proving its reliability. In Fig. 6 the model well reproduces the influence of the Gould Belt:
the two reddening maxima at $l\approx0^{\circ}$ and $155^{\circ}$ predominantly to the north and the south of the Galactic
plane, respectively. This figure can be compared with Fig. 2 in Gontcharov (2009b).

\begin{figure*}
\includegraphics{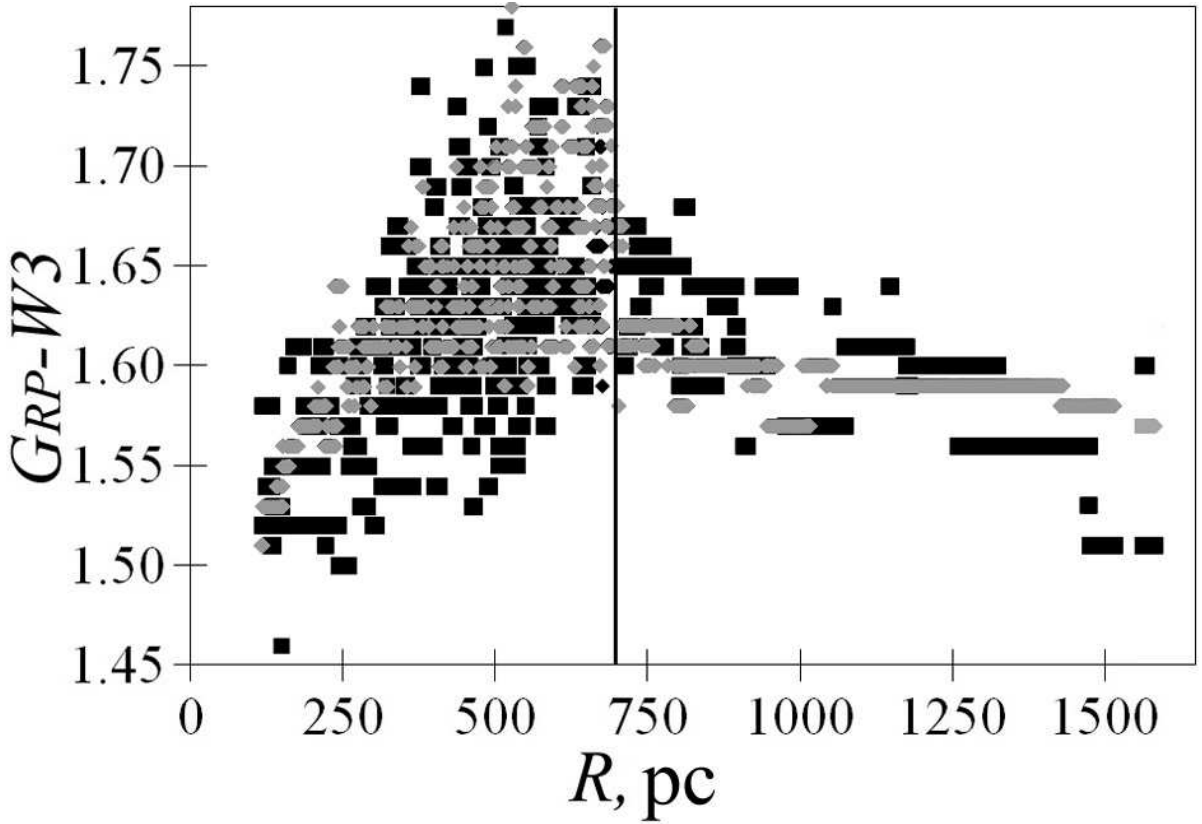}
\caption{$Mode(G_\mathrm{RP}-W3)$ versus $R$: the initial data and the model are indicated by the black squares and gray diamonds,
respectively. The vertical straight line marks $R=700$ pc.
}
\label{rgrw3}
\end{figure*}

The standard deviation of the initial $G_\mathrm{RP}-W3$ is 0.22 mag for individual stars and 0.13 mag for the mode of the color 
in cells of space containing 400 stars each. The solution presented in Table 1 gives a standard deviation of the residuals of 0.186
and 0.087 mag for individual stars and the modes of the color, respectively. For comparison, the analogous standard deviations 
of the residuals when modeling the same data by one layer with a barometric dust distribution without any dependence of the
reddening on longitude turned out to be noticeably larger, 0.2 and 0.1 mag, respectively. Thus, the solution obtained explains 
best much of the scatter of observed $G_\mathrm{RP}-W3$. This can also be seen from Fig. 3.

As has been noted previously, the overwhelming majority of sample stars are rather close to the Galactic plane. Therefore, the 
medians of the age and metallicity predicted by TRILEGAL for the sample stars hardly differ from those for the stars near the 
Galactic plane, 2 Gyr and $[Fe/H]=-0.1$, if we adopt the gradients of age from Casagrande et al. (2016) and metallicity from Soubiran et al. (2008) and Gontcharov (2016a). The theoretical PARSEC and
MIST isochrones for 2 Gyr and $[Fe/H]=-0.1$ are shown in Fig. 3b against a background of the HR diagram for the sample under 
consideration after the correction for the reddening $E(G_\mathrm{RP}-W3)$ and extinction $A_\mathrm{W3}$ found from the CCM89 
extinction law. We used the PARSEC~1.2S$+$COLIBRI~S35 version with options by default, including the mass loss on the giant branch 
$\eta_{Reimers}=0.2$, and the MIST 1.2 version with the rotation of stars with a velocity of 0.4 of the critical one. 
Gontcharov and Mosenkov (2017a) showed that rotation does not affect the isochrones in the giant clump region at all. 
We see that the mode of the distribution of stars agrees equally well with the PARSEC and MIST isochrones, given that
the disagreement between the isochrones can be attributed to an imperfection of the isochrones at a level of a few hundredths of 
a magnitude (for a discussion, see Gontcharov et al. 2019).

\section*{DISCUSSION}

Note some features of the spatial dust distribution that follow from the solution found. As expected, the direction of maximum 
reddening in the equatorial layer (with $l=90^{\circ}-\Phi=50^{\circ}$) falls into the first Galactic quadrant; the Gould Belt 
is oriented in such a way that its semimajor axis (with $l\approx\phi-\theta+\eta-90^{\circ}=43^{\circ}$) falls into the first 
Galactic quadrant; the direction of maximum reddening in the Gould Belt (with $l\approx\phi-\theta-90^{\circ}=15^{\circ}$) 
roughly corresponds to the direction toward the big Scorpius--Ophiuchus--Sagittarius cloud complex. Furthermore, the high
eccentricity (e = 0.95) of the ellipse of the Gould Belt midplane suggests that the bulk of the dust in the Belt is contained 
only in two regions with $-30^{\circ}<l<+30^{\circ}$ and $135^{\circ}<l<180^{\circ}$ (as can also be seen from Fig. 6).

The derived position of the Gould Belt center in the second quadrant ($-x_0=-20$, $-y_0=100$, $-z_0=-7$ pc) at a distance of 
102 pc from the Sun slightly differs from the previously deduced values (for a review, see Bobylev 2014). As has been noted 
previously, the value of $x_0$ found is possibly shifted relative to the true one due to the larger reddening toward the
Galactic center. In addition, the dust distribution in the Belt can differ from the distribution of young stars, from which the 
Sun’s position relative to the Belt center is usually determined. For the same reason, the relatively large scale heights of 
the dust layers found in this study (or the scale height of a single dust layer in those regions where the layers are poorly 
separated) can also differ from the universally accepted scale heights for young Galactic objects presented, for example, by 
Gontcharov (2012c) and Bobylev and Bajkova (2016a, 2016b). These estimates are approximately twice those in the previous solutions
based on the model under consideration. Apparently, the thickness of the layer (layers) was previously underestimated
due to the shortage of stars with large $|Z|$. The estimate obtained here is more preferable, because it is based on a sample 
of stars that is much more extended in the direction of the $Z$ axis and definitely spans the entire thickness of the dust layer
(layers) for the first time. It is important that the values found are consistent with the direct estimate by Kos et al. (2014) 
for long-wavelength radiation and coarse dust mentioned in the Introduction and, thus, confirm the hypothesis outlined in the 
Introduction that larger dust grains are encountered farther from the Galactic plane.

This hypothesis is also confirmed by the rather large mean or median reddening $E(G_\mathrm{RP}-W3)=0.16\pm0.02$ mag found in 
this study for regions with $|Z|>600$ pc (i.e., through the entire dust half-layer above and below the Sun). With the CCM89 
extinction law this gives $E(B-V)=0.08\pm0.01$, which exceeds appreciably the most probable estimate in the range 
$0.04<E(B-V)<0.06$ (Gontcharov and Mosenkov 2018). From one color considered here we cannot draw any conclusions about the actual
extinction law and its spatial variations. In future we are going to use multiband photometry for this purpose. However, to 
explain this contradiction, we can admit an extinction law with $A_\mathrm{G_{RP}}=2.67E(B-V)$ at high latitudes, which is 
provided, for example, by the CCM89 extinction law with $A_V/E(B-V)=3.9$ instead of 3.1, to obtain an estimate of $E(B-V)=0.06$ 
through the Galactic dust half-layer.

\begin{figure*}
\includegraphics{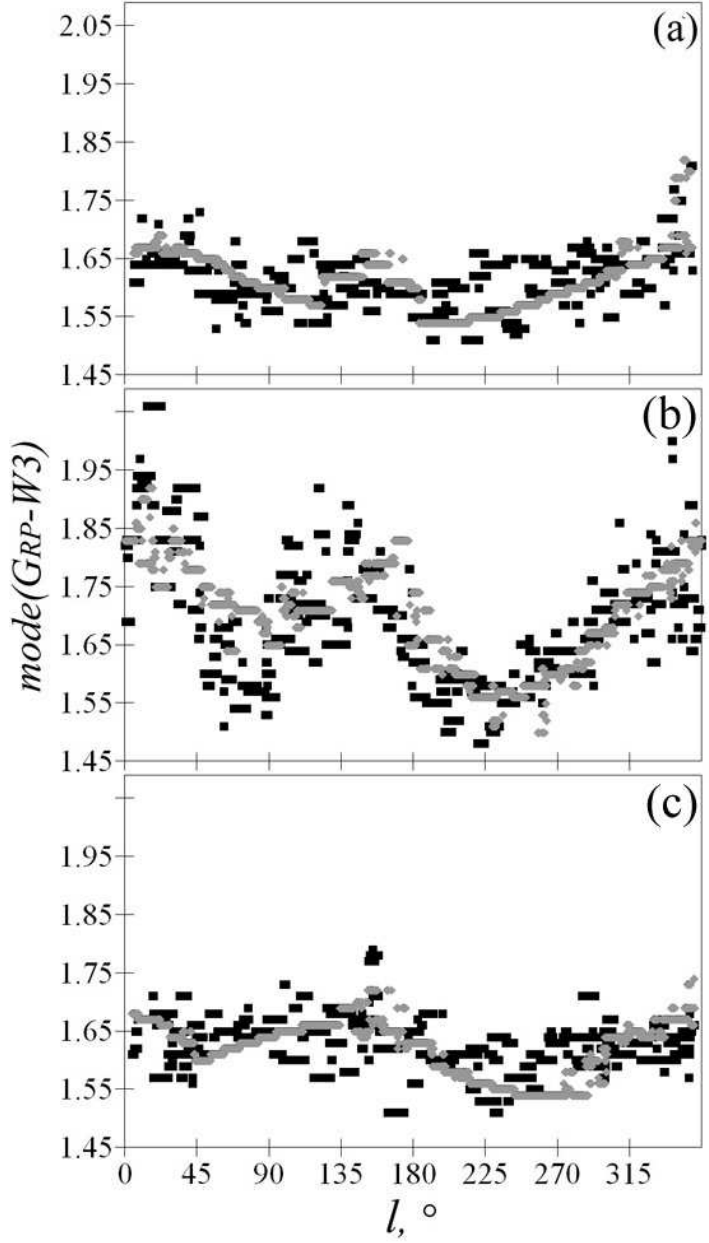}
\caption{$Mode(G_\mathrm{RP}-W3)$ versus Galactic longitude $l$ for the layers $Z>140$ pc (a), $-140<Z<140$ pc (b),
and $Z<-140$ pc (c). The initial data and the model are indicated by the black squares and gray diamonds, respectively.
}
\label{long}
\end{figure*}

\section*{FLUCTUATIONS OF THE DUST MEDIUM}

The observed scatter of $G_\mathrm{RP}-W3$ colors for the stars of our sample includes the following: (1) the scatter due to the 
reddening predicted by the model, including the scatter due to the change in $(G_\mathrm{RP}-W3)_0$ with $|Z|$; (2) the scatter 
due to the small-scale deviations of the reddening from the one predicted by the model (i.e., the fluctuations of the dust medium
across the line of sight mentioned in the Introduction); (3) the natural scatter of color for clump giants; and (4) the natural 
scatter of colors for branch giants. Thus, the derived residuals include the scatter from items (2)–(4). The present-day 
Galaxy models allow us to estimate the dereddened color $(G_\mathrm{RP}-W3)_0$ and its scatter for the stars of our sample as 
a complete sample of giants with a certain color in a certain space. For our sample TRILEGAL gives 80 and 20\%
of the clump and branch giants, respectively, while their distribution in color is well fitted by Gaussians with means and 
standard deviations of $1.42\pm0.1$ and $1.55\pm0.1$, respectively. If we take into account these Gaussians together with the 
scatter due to the reddening from the model, then the residual scatter is indicative of the expected small-scale deviations
of the individual stellar reddenings from the values predicted by the model.

Figure 7 shows the standard deviations $\sigma(G_\mathrm{RP}-W3)$ of the initial data (thick black curve), the reddening
from the model (thin black curve), and the residual fluctuations after allowance for the model and the dispersion of the dereddened 
color for the stars (gray curve) as a function of (a) $R$, (b) $l$ for the layer $|Z|<140$ pc, and (c) $Z$. As expected, the model 
gives a smoothed estimate for the fluctuations of the medium, describing only some of them. The fluctuations of the medium are 
especially large near the Galactic plane ($|Z|<350$ pc) at $R>200$ pc. Near the Sun and far from the Galactic plane the model in 
combination of the natural scatter of dereddened stellar colors explain the entire or almost entire observed scatter of giants
in color.

\begin{figure*}
\includegraphics{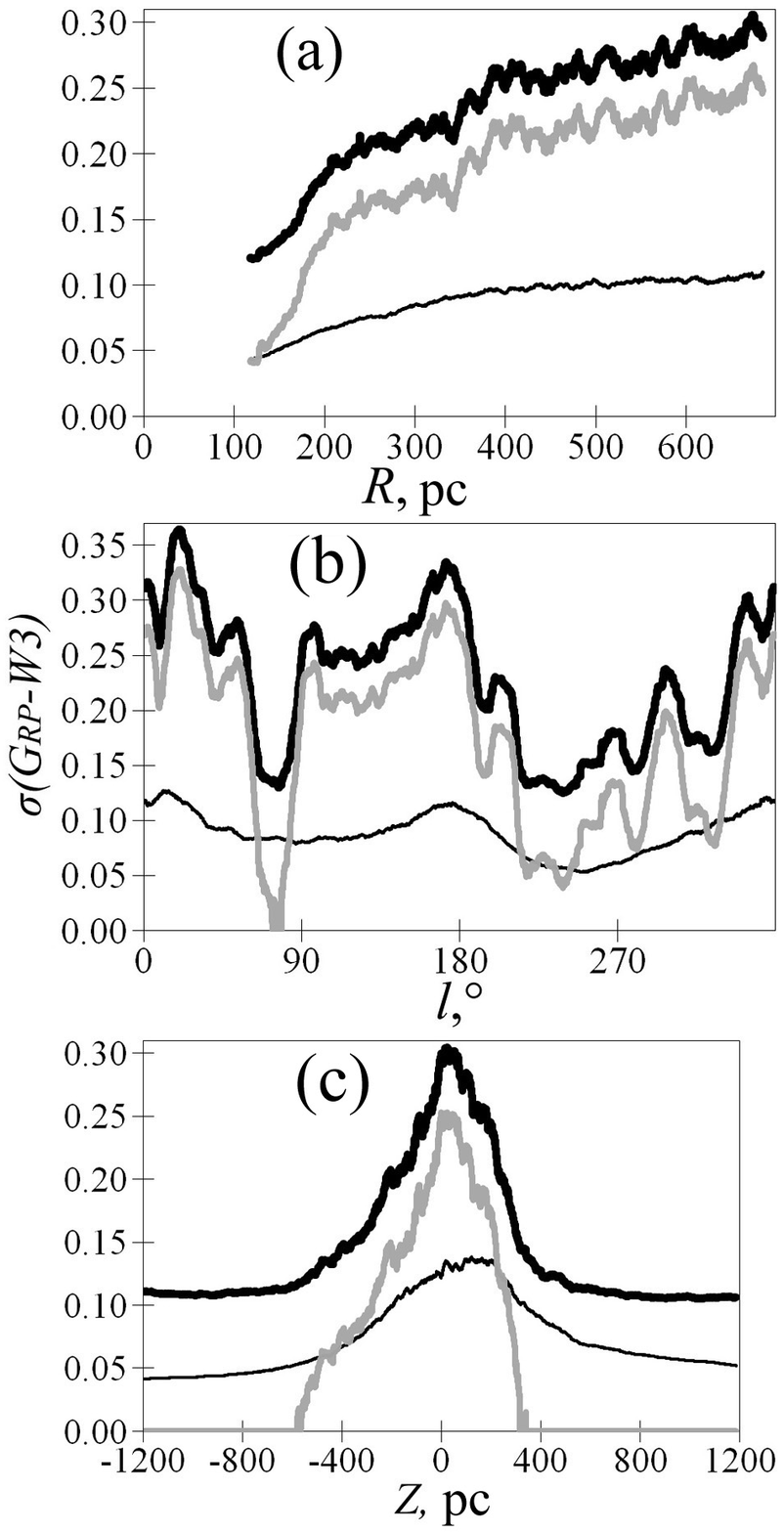}
\caption{Standard deviation $\sigma(G_\mathrm{RP}-W3)$ of the initial data (thick black curve), the reddening from the model
(thin black curve), and the residual fluctuations after allowance for the model and the dispersion of the dereddened
color for the stars (gray curve) versus $R$ (a), $l$ (b) for the layer $|Z|<140$ pc, and $Z$ (c).
}
\label{disp}
\end{figure*}

\begin{figure*}
\includegraphics{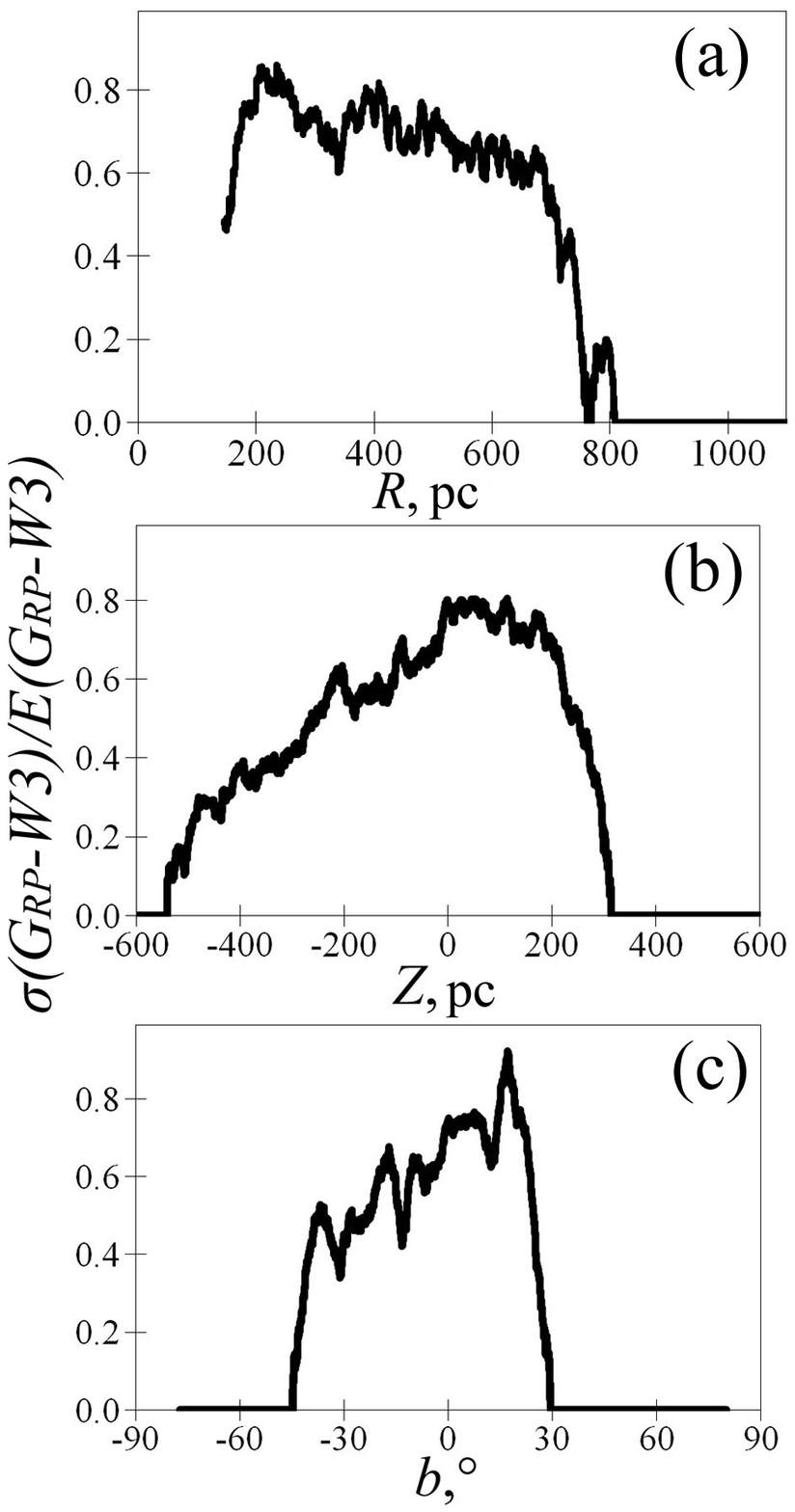}
\caption{Ratio of the standard deviation of the residual fluctuations $\sigma(G_\mathrm{RP}-W3)$ to the median of the derived
reddening $E(G_\mathrm{RP}-W3)$ versus $R$ (a), $Z$ (b), and $b$ (c).
}
\label{ratio}
\end{figure*}

Figure 8 shows the ratio of the standard deviation of the residual fluctuations $\sigma(G_\mathrm{RP}-W3)$ (after allowance for 
the model and the natural scatter of dereddened colors) to the median of the reddening $E(G_\mathrm{RP}-W3)$ found from the 
model as a function of (a) $R$, (b) $Z$, and (c) $b$. This ratio is indicative of a random relative error in the estimates from the
model. We see that this quantity is close to 80\% near the Galactic plane far from the Sun and drops to 20\% or even less at 
$|Z|>400$ pc (the sharp drop at $R>700$ pc is explained by the cylindrical shape of the space under consideration). The larger value
to the north than to the south of the Galactic plane is explained by the orientation of the Gould Belt (it rises to the north of 
the plane at the longitudes of the Galactic center) in combination with the general growth of the fluctuations toward the Galactic
center. Naturally, this ratio actually cannot drop to zero, i.e., the model cannot give perfectly accurate predictions. 
The model imperfection manifests itself as an asymmetry of the residual between the northern and southern hemispheres, as nonzero 
mean residuals in some regions of space, etc. As a result, we can estimate that near the Sun and at high latitudes the model 
predicts the reddening $E(G_\mathrm{RP}-W3)$ with an accuracy of 0.02 mag or the reddening $E(B-V)$ with an accuracy of 0.01 mag 
in accordance with the CCM89 extinction law. This is a rather high accuracy compared to the typical error for other reddening
maps and models at high latitudes, $\sigma(E(B-V))>0.02$ (Gontcharov and Mosenkov 2019). In general, the proposed model gives 
estimates comparable in accuracy or more accurate in random terms than do other maps and models in regions where the ratio
considered in Fig. 8 is less than 50\%. We see that these are regions with R < 200 pc and |$|b|>25^{\circ}$.
However, the worst model predictions at 80\% fluctuations, for example, for the median of the reddening in the space under 
consideration, appear as $E(G_\mathrm{RP}-W3)=0.22\pm0.18$. For such a balance the direct reddening estimates for a specific 
star based on its photometry or spectroscopy are more preferable.

\section*{CONCLUSIONS}

In this study we calculated the parameters of our new complicated version of the three-dimensional model for the spatial dust 
distribution in Galactic solar neighborhoods to refine the properties of the Galactic dust layer. A sample of 93\,992 clump giants
with a small admixture of branch giants with accurate parallaxes and photometry from Gaia DR2 was used for this purpose. 
This sample is complete in a spatial cylinder with a radius of 700 pc around the Sun extending to $|Z|<1800$ pc. Thus, this is the
first sample of stars in astronomy that is complete in the entire space across the dust layer in Galactic solar neighborhoods.

The spatial $G_\mathrm{RP}-W3$ color variations of the sample stars based on their photometry in the Gaia DR2
$G_\mathrm{RP}$ and WISE $W3$ bands are explained by the reddening of the stars and the linear change of their dereddened color with $|Z|$. 
This allowed 19 parameters of the model proposed by Gontcharov (2009b) to be calculated. As in the previous version, the model
suggests two dust layers, along the Galactic equator and in the Gould Belt, that intersect near the Sun at an angle of 18$^{\circ}$. 
In contrast to the previous version, the new version of the model treats the Gould Belt layer as an ellipse with a semimajor axis 
of 600 pc and an eccentricity of 0.95 decentered relative to the Sun. A scale height of $170\pm40$ pc was found for both layers.

A rather large reddening $E(G_\mathrm{RP}-W3)=0.16\pm0.02$ through half of the Galactic dust layer was found for giants far from 
the Galactic plane ($|Z|>600$ pc, i.e., definitely behind the dust half-layers northward and southward of the Sun). This can be 
explained by adopting the extinction law by Cardelli et al. (1989) at high latitudes with $A_\mathrm{V}/E(B-V)=3.9$ instead of 3.1.

The modes of the absolute magnitude $M_\mathrm{W3}=-1.70\pm0.02$ and the dereddened color 
$(G_\mathrm{RP}-W3)_0=(1.43\pm0.01)-(0.020\pm0.007)|Z|$, where $Z$ is expressed in kpc, were calculated for the giant clump
near the Sun. These estimates are consistent with the estimates from the theoretical PARSEC and MIST isochrones for a sample 
dominated by giants with an age of 2 Gyr and metallicity $[Fe/H]=-0.1$ in agreement with the TRILEGAL stellar population model.

A comparison of the dispersions of the observed color for the sample stars, the reddening from the model, and the dereddened color 
of clump and branch giants allowed the residual dispersion to be considered as natural small-scale density fluctuations of the
dust medium relative to the mean reddening calculated from the model. Because of these fluctuations, the reddening of a specific 
star at a certain point of space can differ from the model reddening by a random value that decreases from 80 to $<$20\% of the
model reddening when passing from low latitudes far from the Sun to the remaining space. This means that near the Sun 
($R<200$ pc) and at middle and high latitudes ($|b|>25^{\circ}$) the proposed model is at the accuracy level of the best 
reddening maps and models or surpasses them. In the remaining space the model, along with other maps and models, cannot
take into account the large (in amplitude) small-scale fluctuations of the dust medium and, thus, is inferior to the direct 
reddening measurements for a specific star based on its photometry and spectroscopy.

\section*{ACKNOWLEDGMENTS}

I thank the referees for their useful remarks. I also thank A. Dryanichkin who provided the computer resources. The resources of 
the Strasbourg Astronomical Data Center (http://cdsweb.u-strasbg.fr), including the SIMBAD database and the X-Match
service, the TRILEGAL service (http://stev.oapd.inaf.it/cgi-bin/trilegal), and the PARSEC (http://stev.oapd.inaf.it/cmd) and MIST
(http://waps.cfa.harvard.edu/MIST/) isochrone services were used in this study. This publication makes use of data products from the Wide-field Infrared Survey Explorer, which is a joint project of
the University of California, Los Angeles, and the Jet Propulsion Laboratory/California Institute of Technology.
This work has made use of data from the European Space Agency (ESA) mission Gaia (https://www.cosmos.esa.int/gaia), 
processed by the Gaia Data Processing and Analysis Consortium (DPAC, https://www.cosmos.esa.int/web/gaia/dpac/consortium).


\newpage

\end{document}